# Abstract Machine as a Model of Content Management

Sergey V. Zykov
ITERA Oil and Gas Company
Moscow, Russia
e-mail: szykov@itera.ru

**Abstract**[1]

Enterprise content management is an urgent issue of current scientific and practical activities in software design and implementation. However, papers known as yet give insufficient coverage of theoretical background of the software in question. The paper gives an attempt of building a state-based model of content management. In accordance with the theoretical principles outlined, a content management information system (CMIS) has been implemented in a large international group of companies.

## 1. Introduction

Under the current rapid market development with company merges and acquisitions, the issue of creating a uniform enterprise information infrastructure is becoming extremely urgent.

According to statistical estimates, up to 80% of the information used in mission-critical projects of large companies, is stored beyond the major information systems. Such information (which is often weak-structured or unstructured) makes the enterprise content.

Thus, it becomes evident that design and implementation of enterprise content management software is a must for a company to get advantage under the stiff competition of present-day business.

The objectives of the papers are:

1. analysis of existing approaches to modeling (enterprise) content management;
2. construction of a formal model of content management information system (CMIS);
3. outlining design and implementation scheme of the information system of the kind;
4. overview of CMIS implementation results in a large corporation and recommendations of the software improvement.

## 2. Theoretical Background

An innovative and creative synthesis of typed lambda calculus, combinatory logic, category theory and theory of computation [15] is suggested.

Therewith, the CMIS formal model is built in a form of an abstract machine that operates in a domain of Cartesian closed categories (c.c.c.), where identity and composition operations are defined [3].

## 3. Related Papers

In [1] an exhaustive description of lambda calculus [2,8] is given, which is a formal notation for functional programming languages modeling [4,10,13].

In [11,12,15,17] the *denotational*, *domain*-based approach to semantics is presented (let us informally refer to domains as to a special kind of sets). The approach assumes analysis of syntactically correct programming language constructs (or *denotates*) from the viewpoint of the possibility to evaluate them by means of special functions.

A practical achievement in the field has been programming language semantics formalization in a form of an abstract machine [10,13] that involved such a concept as *state*.

Studies in *operational* semantics of programming languages resulted in *axiomatic* method [9] that models (cause-and-effect) relationships between certain instructions (operators) of programming languages.

Operational semantics development has also resulted in *inductive statement* method [6], used for modeling semantics in program information flows.

Later on, a number of AM formalizations has been implemented, of which a noticeable one is the *categorical abstract machine* (CAM) [3,4].

---

# 4. CMIS Formal Model: a General Scheme

As we have already mentioned above, an abstract machine for content management (AMCM) is a fully adequate solution for CMIS formal model. The AMCM version introduced in the paper has much in common with CAM, named so because it is operating in a domain of c.c.c.

The concept of *state* is the basis for AMCM, as well as of CAM. Therewith, at any given time AMCM is fully characterized by its (certain) state. The process (or, to be more exact, cycle) of AMCM functioning may be formalized by explicit enumeration of all its possible state changes. Thus, AMCM *dynamics* is formalized.

AMCM is based on fundamental theoretical concepts of finite sequences c.c.c., and computation theory. Strictly speaking, AMCM implements a function that maps portal page templates into portal pages. From the formal model viewpoint, such a mapping actually implies binding (template element) variables with their respective (portal page elements) values, i.e. variable-to-value binding occurs.

Therewith, all of the variables are typed. Thus, a type-checking mechanism is necessary for correct binding, i.e. template element types must be matched with the respective portal page elements. A trivial case of type-checking is direct type by easy-to-implement predicates (*IsNum* for numeric expressions, *IsBool* for Boolean ones etc.). Such a matching is implemented directly for (atomic) types. Aggregate (higher order) functions, constructed by application of elementary ones to each other, are used for composite types type-checking.

A possible basis of describing semantics is D.Scott's theory of semantic domains [15]. Under the approach, types of atomic templates (which are content elements) are contained in standard domains, while types of aggregate templates are built by domain constructors.

AMCM formal semantics is formalized in the following order:

1) *standard* (commonly used within the model) domain definition;
2) *finite* (containing explicitly enumerable elements) domain definition;
3) domain *constructor* (operations of forming new domains out of the existing ones) definition, i.e. definition of ways of building domains;
4) aggregate domains formalization by applying domain constructors to standard and finite domains.

Let us start our formal model construction by defining the domain constructors.

Under a functional space from domain $D_1$ into domain $D_2$ let us imply a domain $[D_1 \rightarrow D_2]$ containing all possible functions with definition range of domain $D_1$ and value range of domain $D_2$:

$$[D_1 \rightarrow D_2] = \{f \mid f: D_1 \rightarrow D_2\}.$$

Under Cartesian (or direct) product of domains $D_1, D_2, \ldots, D_n$ let us imply a domain of all possible n-s of the following form:

$$[D_1 \times D_2 \times \ldots \times D_n] =$$
$$= \{ (d_1 \times d_2 \times \ldots \times d_n) \mid d_1 \in D_1, d_2 \in D_2, \ldots, d_n \in D_n, \ldots \}.$$

Under sequence $D^*$ let us imply a domain of all possible final sequences of the following form:

$$d=(d_1,d_2,\ldots,d_n)$$

containing elements $d_1,d_2,\ldots,d_n,\ldots$ of domain D, where n>0.

Finally, under disjunctive sum let us imply a domain defined as

$$[D_1+ D_2+\ldots+D_n] = \{ (d_i, i) \mid d_i \in D_i, 0<i<n+1 \},$$

where membership of elements to component domains $D_i$ is uniquely identified by $d_i$ functions.

Let the formal AMCM language contain a set of expressions E that includes logical, numerical and text constants, a set if identifiers I, and an assignment command that writes appropriate content to the template slot(s). Let us formalize AMCM language syntax as a BNF:

$$E ::= \text{true} \mid \text{false} \mid 0 \mid 1 \mid I.$$

Let us note that within the model (which is solely used for illustration purposes and thus is a simplified one) the expressions include logical (true and false) and integer (limited to 0 and 1) constants as well as identifier set (I).

The formal AMCM language also contains a command set C, which can be formalized by a BNF as follows:

$$C ::= I=E \mid \text{if (E) C1 else C2} \mid C1;C2$$

Let us note that the command set includes assignment (I = E), condition (if (E) C1 else C2) and command sequence (C1;C2).

AMCM language syntax subdivision into expressions and commands is given for pure illustrative purposes.

Let us construct the formal model of AMCM language semantics according to the formal syntax of the language presented earlier.

First of all we need to define syntax domains (i.e. the domains that characterize major syntax categories) for identifiers (Ide domain), expressions (Exp domain) и and commands (Com domain).

Further, let us present the computational model definition on the basis of syntax domains.

Now, let us define semantic functions (E for Exp domain, C for Com domain etc.) that map syntax constructs of the AMCM language into respective semantic representations.

Finally, let us formulate definitions of semantic statements in terms of AMCM state changes.

During AMCM program execution the state change occurs. State concept includes short-term memory, input and output parameters and values are bound, which, in essence, is analogous to variable binding in lambda calculus. A state can also have an "unbound" value, which means there is no binding of an identifier to a value and is quite similar to a free variable in lambda calculus.

According to the outline, let us turn to syntax domains description, which completely defines the AMCM language syntax:

Ide = {I | I – identifier};

Com = {C | C – command};

Exp = {E | E – expression}.

Let us unite the set of all valid identifiers of AMCM language into Ide domain, the set of all valid identifiers – into Com domain, and, the set of all valid expressions – into Exp domain.

Further, let us formulate the state-based computation model for AMCM language:

State = Memory × Input × Output;           (s)

Memory = Ide → [Value + {unbound}];        (m)

Input = Value*;                             (i)

Output = Value*;                            (o)

Value = Int + Bool + String.                (v)

Let us note that CMIS state is always defined by AMCM memory state. Therewith, under memory mapping from identifier domain to value domain (i.e. similar to variable binding to value in lambda calculus) is implied. In order to correctly handle exceptions that occur in case of free variables, an additional "unbound" domain element is introduced. Domain of values is a disjunctive sum of domains containing Int, Bool and String AMCM language types.

According to our outline, let us proceed to semantic statements describing denotate (i.e. correctly built constructs) values of AMCM language.

Let us summarize semantic statements for AMCM language expressions:

**E** : Exp → [State → [[Value × State] + {error}]];

**E** [E]s = (v,s'),

where

v is a value of expression E in state of s,

s' is the state after binding;

**E** [E]s = error,

if type mismatch error occurs.

The statements presented imply that AMCM language expression evaluation results in such a state change that either variable binding occurs, or (if binding is impossible due to type mismatch of the variable and the value) an error is generated. Therewith, the program state is changed from s to s'.

Let us present (type defining) semantic statement for AMCM language commands:

**C**: Com → [ State → [ State + {error} ] ].

The above statement means that the language command evaluation results in AMCM state change, and a situation is possible (e.g., type mismatch during an assignment), when an error is generated.

According to the outline, let us proceed to semantic statements describing certain AMCM language denotate values. Let us consider semantic statements for AMCM language denotates of integer constants:

**E** [0] s = (0, s);

**E** [1] s = (1, s);

The above statements mean that integer constant denotates are their values (in a form of ordered pairs "value"–"state"), and the program state remains unchanged. Let us consider semantic statements for AMCM language denotates of logical constants:

**E** [true] s = (true, s);

**E** [false] s = (false, s);

As it can be concluded from the statements, denotates of logical constants are the values of the constants (in a form of ordered pairs "value"–"state"), and the program state remains unchanged. Let us consider semantic statement for AMCM language denotates of identifiers:

**E** [I] s = (m, I = unbound) error, → (m, I, s).

As it can be concluded from the statement, if binding is possible, identifier denotates are the identifiers bound to the respective variables (in a form of ordered triples "value in the memory" – "identifier" – "state"), and the program state remains unchanged; otherwise an error message is generated.

Semantic function for assignment command of AMCM has a type of

**C**: Com → State → [State + {error}].

Thus, CMIS template substitution by the content may result in AMCM state change, and in some cases (such as template and content elements type mismatch) – in error generation.

Let us present semantic statement for AMCM command that performs content value assignment to a template element:

$C [I = E] = E [E] * \lambda v (m, i, o) . (m [v/I], i, o).$

As it can be judged from the statement, the new AMCM state is different from the old one by the substitution of content value v for template identifier I in AM memory.

## 5. CMIS Design and Implementation Features

Problem domain peculiarities require assignment-based content management support.

The software interface should provide dynamic variation of input fields, flexible adjustment of (meta)data access rights and consistent (meta)data integrity.

The software architecture should provide interoperability, expandability, flexible adaptation to problem domain changes, and fast and reliable (meta)data updates.

## 6. CMIS Implementation Results, Recommendations and Prospects

The paper presents a state-based abstract machine (AMCM) to formalize electronic content publishing process in an Internet portal.

On the basis of principles discussed in the paper, a CMIS has been designed and implemented that supports an official Internet site and an information Intranet portal of a large international group of companies.

Elaborate study of the model supporting the CMIS implemented is to take extra considerable amounts of time and labor.

The author is going to continue studies of the formal models that support information systems for (enterprise) content management.